\documentclass[twocolumn,showpacs,fleqn,nobibnotes]{revtex4}

\usepackage{amsmath}
\usepackage{graphicx}
\usepackage{float}

\newcommand{\xgrv}{x_{\text{GRV}}}
\newcommand{\cms}[1]{#1^{(R)}}

\begin{document}

\title{DGLAP evolution extends the triple pole pomeron fit}
\author{G. Soyez}
\email{g.soyez@ulg.ac.be}
\affiliation{Inst. de Physique, B\^{a}t. B5, Universit\'{e} de Li\`{e}ge, Sart-Tilman, B4000 LI\`{e}ge, Belgium}

\begin{abstract}
We show that the triple pole pomeron model \cite{CMS} provides an initial condition for a DGLAP evolution \cite{DGLAP} that produces a fit to high $Q^2$ experimental DIS data. We obtain good $\chi^2$ for initial scales down to 3 GeV$^2$. Values of the initial scale smaller than 1.45 GeV$^2$ are ruled out at the 90\% confidence level.
\end{abstract}

\maketitle

\section{Introduction}

We have shown in a previous paper \cite{CMS} that it is possible to fit the experimental data for $F_2^p$ with a double or triple pole pomeron model in the region
\begin{equation}\label{eq:cmsdomain}
\begin{cases}
2\nu \ge 49\:\text{GeV}^2, \\
\cos(\theta_t) = \frac{\sqrt{Q^2}}{2xm_p} \ge \frac{49}{2m_p^2}, \\
Q^2 \le 150\:\text{GeV}^2, \\
x \le 0.3.
\end{cases}
\end{equation}
We have also shown that one can extend the usual $t$-channel unitarity relations \cite{GP} to the case of multiple thresholds and multiple poles. This allowed us to predict $F_2^{\gamma}$ from $F_2^p$ and the $pp$ total cross-section. In the latter case, we have shown that all processes have the same singularity structure. 

However, in the usual parton distribution sets, each parton distribution presents its own singularities. As an example, in the MRST2001 parametrization \cite{MRST2001}, we have
\[
xq(x, Q_0^2) = A (1+B\sqrt{x}+Cx)(1-x)^{\eta_q}x^{\varepsilon_q},
\]
with $\varepsilon_\text{sea}=-0.26$, $\varepsilon_g^{(1)}=-0.33$, $\varepsilon_g^{(2)}=0.09$. In fact, these singularities do not correspond to any singularity present in hadronic cross sections and, conversely, cross section singularities are not present in parton distributions. There must therefore exist a mechanism through which the singularities in partonic distributions disappear and cross section singularities arise when $Q^2$ goes to zero. Such a mechanism is unknown and seems forbidden by Regge theory. In this framework, a singularity structure common both to parton distibutions and to hadronic cross sections is the most natural choice.

At that level, one may ask whether the Regge fit of \cite{CMS} is compatible with pQCD and whether it is possible to have the same singularities in all parton distributions. Actually, although Regge theory \cite{Regge:mz,Regge:1960zc} and DGLAP \cite{DGLAP} evolution both provide well-known descriptions of the structure functions \cite{MRST2001, GRV98, CTEQ6, ZEUSfit, Alekhin, CS, EM, DL}, the connection between the two approaches is unclear. In this paper, we will confront the triple-pole parametrisation, for each parton distribution, and evolve it with DGLAP. This is done by fixing the initial distribution at $Q_0^2$ in order to reproduce the $F_2^p$ value obtained from the global QCD fit \cite{CMS}. We shall see that we are able to produce a fit to experimental data which is compatible both with Regge theory and with the DGLAP equation. This comparison of two aspects of the theory will allow us to split the $F_2$ structure function in smaller contributions and to predict the density of gluons, which is generally not accessible directly from Regge fits. 

Varying the initial scale $Q_0^2$, we can predict where perturbative QCD breaks down. However, due to the application domain \eqref{eq:cmsdomain} of the global fit, the Regge constraint on the initial parton distributions is not valid at large $x$. In order to solve this problem, we use the GRV98 parton distributions \cite{GRV98} at large $x$ ($x>\xgrv$). We shall argue that the results do not significantly depend on the choice of the large $x$ parametrisation. Since we will use leading order (LO) DGLAP evolution, one can choose any of the usual PDF sets to extend our fits to large $x$.

We will show that, within a reasonable region of $Q_0^2$ and $\xgrv$, the triple-pole pomeron model provides an initial condition for LO DGLAP evolution which reproduces the experimental data. The scale $Q_0^2$ should be considered as the minimal scale where perturbative QCD can be applied.

Since a good precision on the gluon density is of primary importance for the LHC, it is also very interesting to look at the prediction of this model for the density of gluons. We will see that the densities we obtain are of the same order of magnitude as in the usual DGLAP fits.

One should mention that such an extension of the triple pole Regge fit by a DGLAP evolution has already been introduced in \cite{Csernai}. However, as we will see, our approach here is different: our parametrisation is much more constrained, we are able to extract a gluon distribution and all the distributions have the same singularity structure. There are also some less important differences in the treatment of the large-$x$ domain.

\section{Perturbative QCD and Regge theory}

\subsection{Perturbative QCD}

In pQCD, the high $Q^2$ behaviour of Deep Inelastic Scattering (DIS) is given by the DGLAP evolution equations \cite{DGLAP}. These equations introduce the {\em parton distribution functions} $q_i(x, Q^2)$, $\bar q_i(x, Q^2)$ and $g(x, Q^2)$, which represent the probability of finding, in the proton, respectively a quark, an anti-quark or a gluon with virtuality less than $Q^2$ and with  longitudinal momentum fraction $x$. When $Q^2 \to \infty$, the $Q^2$ evolution of these densities (at fixed $x$) are given by the DGLAP equations
\begin{eqnarray}\label{eq:DGLAP}
\lefteqn{ Q^2\partial_{Q^2}
\begin{pmatrix}q_i(x,Q^2)\\\bar q_i(x,Q^2)\\g(x,Q^2)\end{pmatrix}}\\
& = &\frac{\alpha_s}{2\pi} \int_x^1 \frac{d\xi}{\xi}
\left.\begin{pmatrix}
 P_{q_iq_j} & . & P_{q_ig}\\
 . & P_{q_iq_j} & P_{q_ig}\\
 P_{gq} & P_{gq} & P_{gg}
\end{pmatrix}\right|_{\frac{x}{\xi}}
\begin{pmatrix} q_j(\xi,Q^2)\\\bar q_j(\xi,Q^2)\\ g(\xi, Q^2) \end{pmatrix},\nonumber
\end{eqnarray}
at leading order. Using these definitions, we have 
\begin{equation}\label{eq:F2}
F_2(x,Q^2) = x \sum_i e_{q_i}^2 \left[q_i(x,Q^2) + \bar q_i(x,Q^2) \right],
\end{equation}
where the sum runs over all quark flavours.

The usual way to use this equation is to choose a set of initial distributions $q_i(x, Q_0^2, \vec{a})$ to compute $q_i(x, Q^2, \vec{a})$ using \eqref{eq:DGLAP} and to adjust the parameters $\vec{a}$ in order to reproduce the experimental data.
This approach have already been successfully applied many times \cite{MRST2001, GRV98, CTEQ6, ZEUSfit, Alekhin} and is often considered a very good test of pQCD.
Nevertheless, these studies do not care about the singularity structure of the initial distributions, ending up with results that disagree with Regge theory, and presumably with QCD.

\subsection{Regge theory}

Beside the predictions of pQCD, we can study DIS through its analytical properties. In Regge theory \cite{Regge:mz,Regge:1960zc}, we consider amplitudes ${\cal A}(j,t)$ in the complex angular momentum space by performing a Sommerfeld-Watson transform. In that formalism, we choose a singularity structure in the $j$-plane for the amplitudes. The residues of the singularities are functions of $t$ and this technique can be applied to the domain $\cos(\theta_t) \gg 1$. For example, we can fit the DIS data or the photon structure function at large $\nu$ (small $x$), and the total cross sections at large $s$. 

The models based on Regge theory \cite{CS,EM,DL,books} use a pomeron term, reproducing the rise of the structure function (cross sections) at small $x$ (at large $s$), and reggeon contributions coming from the exchange of meson trajectories ($a$ and $f$). 
We shall consider here the following parametrisation for the pomeron term
\begin{equation}\label{eq:triple}
a(Q^2)\ln^2\left[\nu/\nu_0(Q^2)\right] + c(Q^2),
\end{equation}
corresponding to a triple pole in $j$-plane \cite{CMS}
\[
\frac{a}{(j-1)^3}-\frac{2a\ln(\nu_0)}{(j-1)^2}+\frac{a\ln^2(\nu_0)+c}{j-1}.
\]
This seems to be the preferred phenomenological choice at $Q^2=0$ \cite{compete}.
Note that the upper expression, given in terms of $\nu$ and $Q^2$, can be rewritten in terms of $Q^2$ and $x=Q^2/(2\nu)$.

In a previous paper \cite{CMS}, we have also shown from unitarity constraints that we can extend the Gribov-Pomeranchuk argument about factorisation of residues to any number of thresholds and to any type of singularities. Hence, if we parametrise the $pp$ and the $\gamma^{(*)}p$ cross sections, we can predict the $\gamma^{(*)}\gamma^{(*)}$ cross-section using the $t$-channel unitarity ($t$CU) relation
\[
A_{\gamma\gamma}(j,Q_1^2,Q_2^2) = \frac{A_{\gamma p}(j,Q_1^2)A_{\gamma p}(j,Q_2^2)}{A_{pp}(j)}+\text{finite terms}
\]
for the amplitudes in the $j$-plane. This relation proves the universality of the singularities, in other words, all singularities present in $\gamma p$ interactions also appear in $\gamma \gamma$ interactions. We have successfully applied the $t$CU rules to the case of double and triple pole pomeron models in the region \eqref{eq:cmsdomain}. Therefore, since our fit keeps consistency with the fits in \cite{CMS} for $Q^2\le Q_0^2$, we also be used to reproduce the $\gamma^{(*)}\gamma^{(*)}$ experimental results for $Q_1^2\le Q_2^2\le Q_0^2$.

\section{Initial distributions}

In our approach it is not possible to dissociate the soft singularity from the perturbative ones, as was done in \cite{DLclose}. However, it is possible to assume that the perturbative essential singularity (at $j=1$) becomes a triple pole at small $Q^2$. This may come from further resummation of pQCD \cite{cgc}. We thus have two regimes: for $Q>Q_0^2$, we have a perturbative DGLAP evolution with an essential singularity, while for $Q^2\le Q_0^2$, the Regge fit applies, and $F_2$ behaves like a triple pole at small $x$.

Due to the fact that the domain \eqref{eq:cmsdomain} does not extend up to $x=1$, we have used the GRV98 \cite{GRV98} parametrisation at large $x$, i.e. for $x > \xgrv$. It is worth mentioning that, in the DGLAP equation \eqref{eq:DGLAP}, the evolution for $x > \xgrv$ does not depend on the distributions below $\xgrv$. This means that the evolution of the GRV98 distribution functions for $x > \xgrv$ is not influenced by the parametrisation we will impose for $x \le \xgrv$.

Since we want to use our fit \eqref{eq:triple} to $F_2$, for $x\le\xgrv$, we want to have an initial distribution of the form ($Q_0^2$ is the scale at which we start the DGLAP evolution)
\begin{equation}\label{eq:initF2}
F_2(x, Q_0^2) = a \log^2(1/x) + b \log(1/x) + c + d x^\eta,
\end{equation}
i.e. described by a triple pole pomeron and an $f$,$a_2$-reggeon trajectory ($\eta=0.31$ as given in \cite{CMS}). Once we have that initial distribution, we can evolve it with DGLAP and compare it with experimental data.

However, the DGLAP equation \eqref{eq:DGLAP} does not allow us to compute $F_2$ directly. Performing linear combinations in \eqref{eq:DGLAP}, one can easily check that the minimal set of densities needed to obtain $F_2$ from the DGLAP equation is given by
\begin{eqnarray}
T      & = & x\left\lbrack (u^+ +c^++t^+)-(d^++s^++b^+)\right\rbrack,\\
\Sigma & = & x\left\lbrack (u^+ +c^++t^+)+(d^++s^++b^+)\right\rbrack,\\
G      & = & x g,
\end{eqnarray}
where $q^+ = q+\bar q$ for $q=u,d,s,c,t,b$.
The evolution equations for these distributions turn out to be
\begin{eqnarray*}
Q^2\partial_{Q^2} T(x,Q^2) & = & \frac{\alpha_s}{2\pi} \int_x^1 \frac{xd\xi}{\xi^2}
P_{qq}\left(\frac{x}{\xi}\right) T(\xi,Q^2),\\
Q^2\partial_{Q^2}
\begin{pmatrix}\Sigma\\G\end{pmatrix} 
  & = &\frac{\alpha_s}{2\pi} \int_x^1 \frac{xd\xi}{\xi^2}
\begin{pmatrix}
 P_{qq} & 2n_fP_{qg}\\
 P_{gq} & P_{gg}
\end{pmatrix}
\begin{pmatrix} \Sigma\\ G \end{pmatrix}
\end{eqnarray*}
and $F_2$ is then given by
\[
F_2=\frac{5\Sigma+3T}{18}.
\]
This clearly shows that, if we want to use \eqref{eq:initF2} as the initial condition for a DGLAP evolution, we need to split it into $T$ and $\Sigma$ contributions, but we also need to introduce a gluon density. In this way, using \eqref{eq:initF2} as the initial condition for the evolution allows us to predict the gluon distribution function.

Since, below $Q_0^2$, we do not use singularities of order larger than 3, we expect this behaviour to be valid for the $T$ and $\Sigma$ distributions. The natural way of separating the initial $F_2$ value given by \eqref{eq:initF2} is thus to consider both $T$ and $\Sigma$ as a sum of a triple pole pomeron and a reggeon. The gluon distribution, being coupled to $\Sigma$, should not contain any singularities either.
Thus, we can write
\begin{eqnarray}
T(x,Q_0^2) & = & a_T \log^2(1/x) + b_T \log(1/x) + c_T + d_T x^\eta,\nonumber\\
\Sigma(x,Q_0^2) & = & a_\Sigma \log^2(1/x) + b_\Sigma \log(1/x) + c_\Sigma + d_\Sigma x^\eta,\nonumber\\
G(x,Q_0^2) & = & a_G \log^2(1/x) + b_G \log(1/x) + c_G + d_G x^\eta.\nonumber
\end{eqnarray}

Most of the 12 parameters in these expressions are constrained. First of all, since the triple-pole pomeron, describing the high-energy interactions, has the vacuum quantum numbers, it will not be sensitive to the quark flavours. This means that, at high energy, one expects $T \to 0$. Therefore, we set $a_T=b_T=c_T=0$. Then, since we connect our parametrisation with GRV's at $\xgrv$, we want the distribution functions to be continuous over the whole $x$ range. Continuity of the $T$ distribution fixes $d_T$ and we finally have
\[
T(x,Q_0^2) = T^{(GRV)}(\xgrv,Q_0^2) \left(\frac{x}{\xgrv}\right)^\eta.
\]

Moreover, we want to fix $F_2(Q_0^2)$ to be equal to $\cms{F_2}$ obtained from our previous global fit (each quantity with a superscript $^{(R)}$ refers to the corresponding quantity obtained from the Regge fit in \cite{CMS}). Since $T$ is entirely known, this constraint fixes all the $\Sigma$ parameters through the relation
\begin{equation}\label{eq:formrel}
\phi_\Sigma = \frac{18\cms{\phi}-3\phi_T}{5},\quad \phi=a,b,c,d.
\end{equation}

At this level, only the gluon distribution parameters are free. However, since the reggeon trajectory is expected to be mainly constituted of quarks, we may exclude its contribution from the gluon density. Thus, we take $d_G=0$. Finally, we used continuity of the gluon density with the GRV distribution at $\xgrv$ to fix $c_G$.

We are finally left with only 2 free parameters: $a_G$ and $b_G$.

Before going to the fits, one must stress that the GRV parametrisation at large $x$ does not modify the triple pole singularity structure of the initial distributions. Actually, one may write the Mellin transform 
\begin{eqnarray*}
\int_0^1 dx\,x^{j-1} q(x) & = & \int_0^{\xgrv} dx\,x^{j-1} q_{\text{regge}}(x)\\
                          & + & \int_{\xgrv}^1 dx\,x^{j-1} q_{\text{grv}}(x).
\end{eqnarray*}
In this expression, the first term generates the triple pole pomeron and the reggeon. The singularities of the second term come from the behaviour near $x=1$. Since parton distributions behave like $x^{\varepsilon-1}(1-x)^n$, the GRV parton distributions give the following contribution
\[
\sum_{k=0}^n (-)^k \begin{pmatrix}n\\k\end{pmatrix} \frac{1-\xgrv^{j+\varepsilon+k-1}}{j+\varepsilon+k-1}.
\]
and the zeroes of the numerator cancel those from the denominator. Thus, using GRV at large $x$ does not interfere with the singularity structure imposed from the low $x$ parametrisation.


\section{Fit}

We will fit the DIS data coming from H1\cite{H1-1,H1-2,H1-3}, ZEUS\cite{ZEUS-1,ZEUS-2}, BCDMS\cite{BCDMS}, E665\cite{E665}, NMC\cite{NMC} and SLAC\cite{SLAC}. We will only consider data for $F_2^p$. We have not included data from $F_2^d$, $F^{\nu N}$, Drell-Yan proccesses and Tevatron Jets for the following reasons
\begin{itemize}
\item for many experiments, most of the data points are at large $x$ or at low $Q^2$. Thus, they do not constrain our fit much.
\item some experiments allow the determination of the valence quark distributions. We do not need them here since we only want the $T$, $\Sigma$ and gluon distributions.
\end{itemize}

\begin{figure}[ht]
\includegraphics{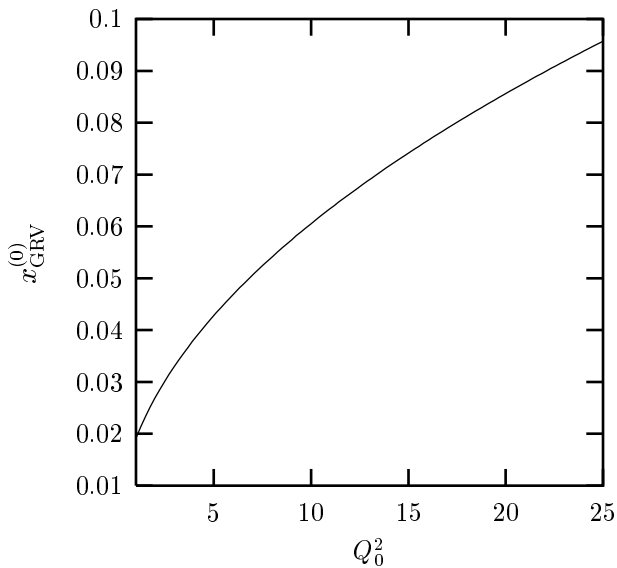}
\caption{Natural value of $\xgrv$ as a function of the scale}
\label{fig:xgrv}
\end{figure}

Since we want to test the domain common to Regge theory and to the DGLAP evolution, we only keep the experimental points verifying
\begin{equation}\label{eq:domain}
\begin{cases}
\cos(\theta_t) \le \frac{49}{2m_p^2},\\
Q_0^2 \le Q^2 \le 3000\:\text{GeV}^2,\\
x\le \xgrv.
\end{cases}
\end{equation}
We have tried several values of the initial scale $Q_0^2$ around 5 GeV$^2$. Given an initial scale, the Regge limit on $\cos(\theta_t)$ translates into a natural value for $\xgrv$
\begin{equation}
\xgrv^{(0)} = \frac{m_p\sqrt{Q_0^2}}{49}.
\end{equation}
A graph of that limit is presented in Fig. \ref{fig:xgrv}.
\begin{figure}[ht]
\includegraphics[scale=0.65]{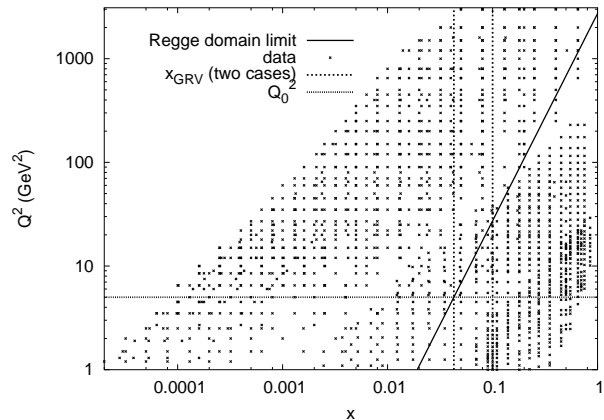}
\caption{Experimental points, Regge domain limit and fit domain limits for $Q_0^2=5$ GeV$^2$ and $\xgrv=\xgrv^{(0)}$ or $0.1$. It clearly appears that without extrapolation, we miss the high-$Q^2$ points.}
\label{fig:pts}
\end{figure}
However, as one can see from Fig. \ref{fig:pts}, if we take that limit on $x$, we cut most of the high $Q^2$ experimental points which are at large $x$. It is therefore interesting to extrapolate the initial distributions to larger $x$, and we have tried some higher values for $\xgrv$.

\section{Results}

\begin{figure}[ht]
\includegraphics{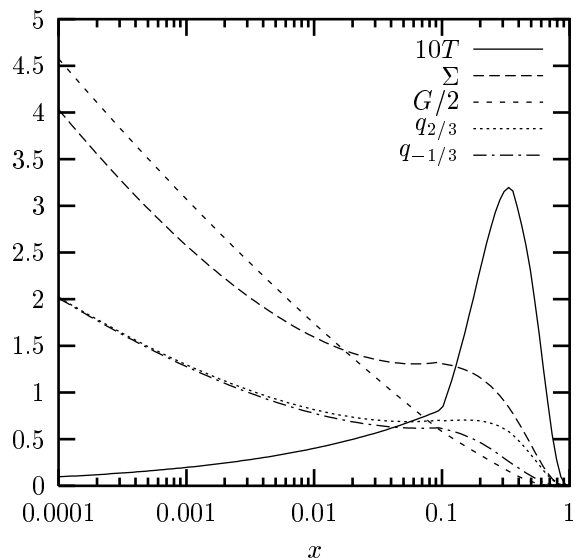}
\caption{Initial distributions for $Q_0^2=5$ GeV$^2$ and $\xgrv = 0.1$. $q_{2/3} = x (u^+ + c^+ + t^+)$ and $q_{-1/3} = x(d^++s^++b^+)$}
\label{fig:distrib}
\end{figure}

\begin{figure}[ht]
\includegraphics{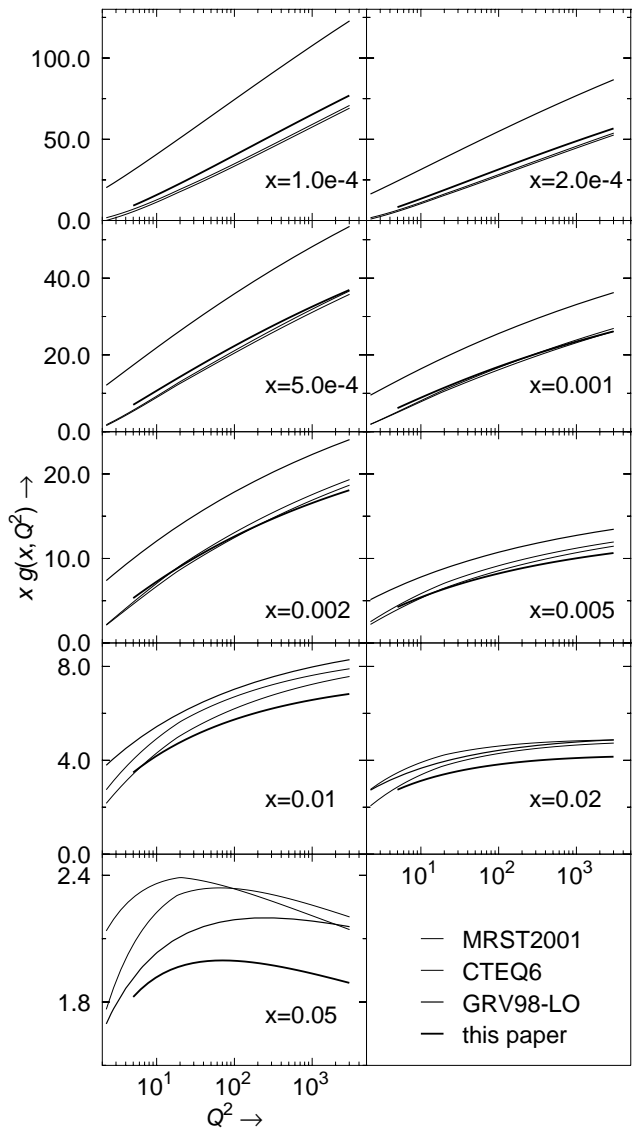}
\caption {Fitted gluon distribution compared with some well known parton distributions}
\label{fig:gluons}
\end{figure}

The results of the fits are given in table \ref{tab:chi2} as a function of $Q_0^2$ and $\xgrv$. We can see that this 2-parameters fit reproduces very well the experimental points in \eqref{eq:domain} for $Q_0^2\ge 3$ GeV$^2$ and $\xgrv \le 0.1$. The values of the fitted parameters, as well as the constrained parameters are given in table \ref{tab:param}.

\begin{table}
\begin{tabular}{|l||c|c|c||c|c|c||c|c|c|}
\hline
\multicolumn{1}{|r||}{$\xgrv$} & \multicolumn{3}{c||}{$\xgrv^{(0)}$} & \multicolumn{3}{c||}{0.1} & \multicolumn{3}{c|}{0.2} \\
\hline
$Q_0^2$ & $\chi^2$ & $n$ & $\chi^2/n$ & $\chi^2$ & $n$ & $\chi^2/n$ & $\chi^2$ & $n$ & $\chi^2/n$ \\
\hline\hline
10.0 & 484 & 515 & 0.939 & 561 & 581 & 0.966 & 774 & 639 & 1.212 \\
\hline
5.0  & 557 & 577 & 0.966 & 676 & 686 & 0.985 & 862 & 744 & 1.159 \\
\hline
3.0  & 633 & 591 & 1.071 & 741 & 735 & 1.008 &    -    &  -  &   -   \\
\hline
\end{tabular}
\caption{$\chi^2$ for various values of $Q_0^2$ and $\xgrv$. ($n$ is the number of experimental point satisfying \eqref{eq:domain})}
\label{tab:chi2}
\end{table}

\begin{table}
\begin{tabular}{|l||c|c||c|c||c|c|}
\hline
$Q_0^2$       & \multicolumn{2}{c||}{3.0}     & \multicolumn{2}{c||}{5.0}     & \multicolumn{2}{c|}{10.0}    \\
\hline
\hline
$a_{\gamma p}$& \multicolumn{2}{c||}{0.00541}& \multicolumn{2}{c||}{0.00644}& \multicolumn{2}{c|}{0.00780}\\
$b_{\gamma p}$& \multicolumn{2}{c||}{0.0712} & \multicolumn{2}{c||}{0.0990} & \multicolumn{2}{c|}{0.142}  \\
$c_{\gamma p}$& \multicolumn{2}{c||}{0.00541} & \multicolumn{2}{c||}{0.0064} & \multicolumn{2}{c|}{0.00780} \\
$d_{\gamma p}$& \multicolumn{2}{c||}{0.890}   & \multicolumn{2}{c||}{1.06}   & \multicolumn{2}{c|}{1.27}   \\
\hline
\hline
$\xgrv$       & $\xgrv^{(0)}$ &     0.1      & $\xgrv^{(0)}$ &     0.1      & $\xgrv^{(0)}$ &     0.1      \\
\hline
\hline
$d_T$         &   -0.0722    &    0.167    &   -0.0478    &    0.166    &    0.0101    &    0.165    \\
$a_G$  &\textbf{0.147}&\textbf{0.00617}&\textbf{0.0908}&\textbf{0.0271}&\textbf{0.158}&\textbf{0.131}\\
$b_G$  & \textbf{-0.852}&\textbf{0.718}&\textbf{0.193} &\textbf{0.822} &\textbf{0.178}&\textbf{0.419}\\
$c_G$         &    3.45      &   -0.495    &    0.595     &   -0.851    &    0.0299    &   -0.463    \\
\hline
\end{tabular}
\caption{Values of the parameters for $\le Q_0^2 \le 10$ GeV$^2$ and $\xgrv\le 0.1$. Only $a_G$ and $b_G$ are fitted, while the other parameters are constrained.}
\label{tab:param}
\end{table}

\begin{figure*}[ht]
\includegraphics{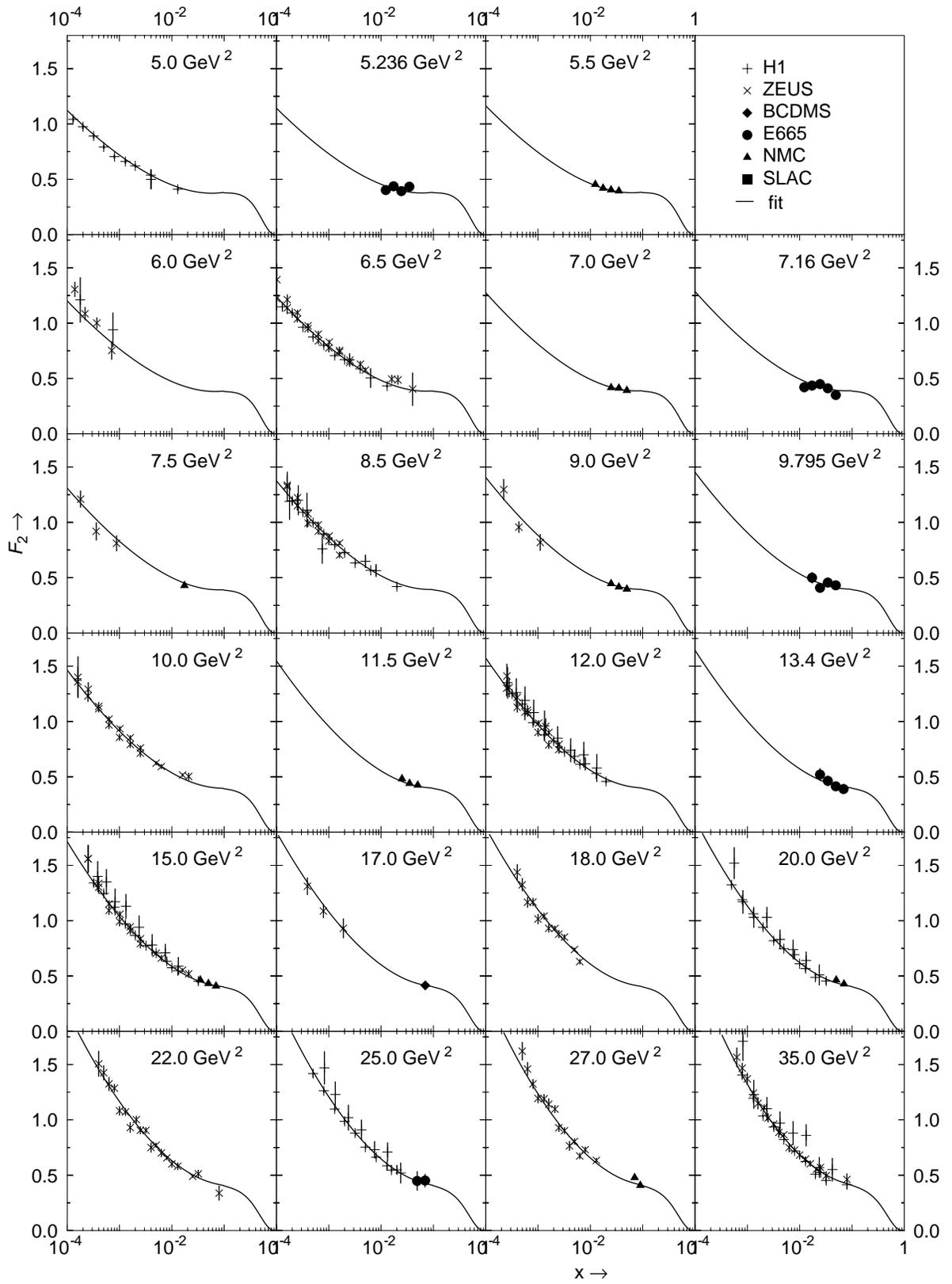}
\caption{$F_2^p$ fit for $Q_0^2=5$ GeV$^2$ and $\xgrv \le 0.1$ (low $Q^2$ values).}
\label{fig:F2fit-low}
\end{figure*}

\begin{figure*}[ht]
\includegraphics{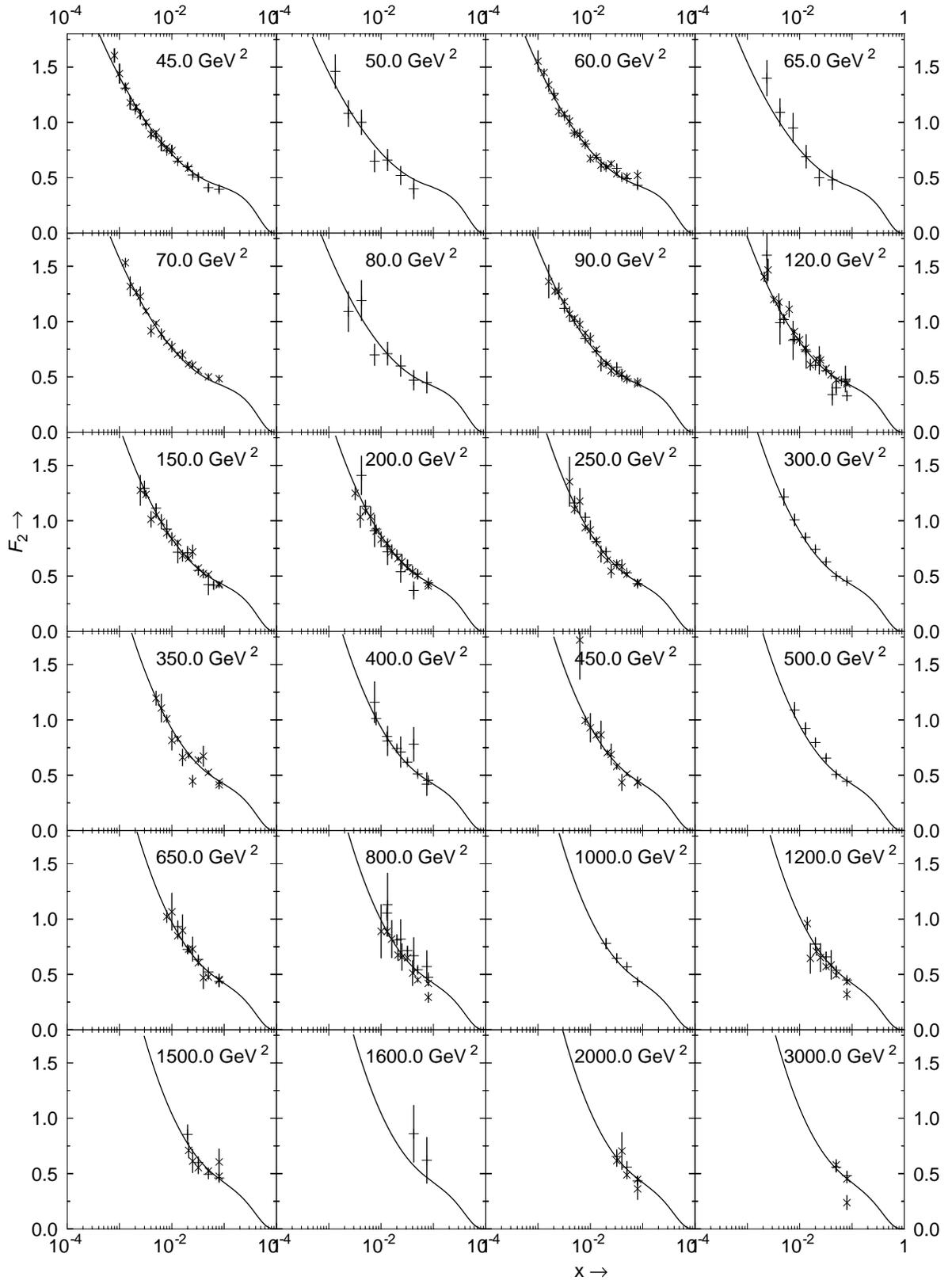}
\caption{$F_2^p$ fit for $Q_0^2=5$ GeV$^2$ and $\xgrv \le 0.1$ (high $Q^2$ values).}
\label{fig:F2fit-high}
\end{figure*}

We show the initial distributions and the $F_2^p$ plot for $Q_0^2=5$ GeV$^2$ and $\xgrv = 0.1$ in Fig. \ref{fig:distrib} and Figs. \ref{fig:F2fit-low},\ref{fig:F2fit-high} respectively.


In Fig. \ref{fig:gluons}, we have compared the gluon distribution obtained from our fit with that of the well known DGLAP fits like GRV\cite{GRV98}, CTEQ\cite{CTEQ6} and MRST\cite{MRST2001}.
One can see that our gluon distribution is of the same order of magnitude as that from other DGLAP fits.

It is also interesting to check whether our results depend on the choice of the large $x$ parametrisation. Since the DGLAP evolution equation couples the small $x$ distributions to the large $x$ ones, at first sight, our results may depend on such a choice. However, looking at the studies of the PDF uncertainties, it can be seen that the large $x$ behaviour of the $T$ and $\Sigma$ distributions hardly depends on the chosen fit down to $x\approx 0.1$. Moreover, in the large $x$ limit, the splitting matrix can be written
\[
\begin{pmatrix} P_{qq} & P_{qg} \\ P_{gq} & P_{gg} \end{pmatrix}
\approx \frac{1}{(1-x)_+}\begin{pmatrix} 2C_F & . \\ . & 2C_A \end{pmatrix}.
\]
Thus, in the large $x$ region, the gluon distribution and the sea are not coupled. Since, in our method, both $T$ and $\Sigma$ are fixed, we study the influence of the gluon distribution on $F_2$. Due to the fact that these are not coupled at large $x$, we expect that our fit does not depend on the large $x$ behaviour of the distributions.

Furthermore, one can see that the $\chi^2$ of the fit remains of order 1 for $0.04\lesssim \xgrv \lesssim 0.15$ and grows when we take $\xgrv \sim 0.01$ or smaller. At that point , parton distributions depend on the chosen parametrisation and the one we used, namely GRV98, does not take into account the latest HERA points. If we want to go to smaller values of $\xgrv$, we need a more recent parametrisation and thus a NLO study. Note that the interval on $\xgrv$ for which we have an acceptable $\chi^2$ hardly depends on $Q_0^2$ for $Q_0^2$ in $[3, 15]$ GeV$^2$, and that the $\chi^2$ of the fit does not change very much in that domain.

Unfortunately, it is quite hard to determine a unique scale $Q_0^2$ or $\xgrv$ from the fit. From Table \ref{tab:chi2}, it is clear that $\xgrv$ can be taken to be 0.1 but can not be pushed up to 0.2. But, as we have argued, for such values of $Q_0^2$ and such high $\xgrv$, we are outside the domain \eqref{eq:cmsdomain} and we may not ensure that Regge theory will still be valid at $x=0.1$ and $Q^2=Q_0^2$. We can thus adopt two different points of view:
\begin{enumerate}
\item we stay in the domain \eqref{eq:cmsdomain}. We have thus $\xgrv = \xgrv^{(0)}$ and we can take $Q_0^2$ down to 3 GeV$^2$. The problem is that as $Q_0^2$ goes down, $\xgrv$ goes down too. And, since high $Q^2$ experimental points have large $x$ values, we do not test $pQCD$ over a large range of $Q^2$ values. This effect can be clearly seen in Fig. \ref{fig:pts} where we have plotted the experimental points, the Regge domain limit and the fit domain for $Q_0^2=5$ GeV$^2$ and $\xgrv=\xgrv^{(0)}$ or $0.1$. It is therefore difficult to predict a ``best value'' for $Q_0^2$.
\item we extrapolate the Regge fit outside the domain~\eqref{eq:cmsdomain}. The amount of points concerned by this interpolation can be seen in Fig. \ref{fig:pts}. In such a case, depending on our confidence in this extrapolation, we can consider that pQCD applies down to 3 GeV$^2$ or 5 GeV$^2$ and $\xgrv\approx 0.1$. This value is compatible with the HERA predictions as well as with the Donnachie-Landshoff prediction \cite{DLclose}. Below 2~GeV$^2$, the $\chi^2$ is larger than 1 whatever $\xgrv$ is and values of the initial scale smaller than 1.45~GeV$^2$ are ruled out at the 90\% confidence level.
\end{enumerate}

\section{Conclusion}
In this paper, we have shown that it is possible to use a very simple analytic form, namely a triple-pole pomeron and a reggeon, as an initial condition for the DGLAP evolution. Applying the constraint from a global QCD fit obtained in a previous paper \cite{CMS} as well as some expected properties of the parton distribution functions, we have shown that we can fit the DIS data in the domain $Q_0^2~\ge~3$~GeV~$^2$, $x\le 0.1$ and $\cos(\theta_t)\ge 49/(2m_p^2)$. This fit has only 2 free parameters in the gluon distribution. 

Our fit is at the interplay between Regge theory and pQCD. We have thus proven that Regge theory can be used to extend a pQCD evolution down to the non-perturbative domain. From the fit, we can also say that the scale down to which we can apply pQCD is of the order of 3-5 GeV$^2$.

Moreover, we have seen that our approach can be used to split $F_2$ in $T$ and $\Sigma$-components with precise physical properties. In this way, it is of prime importance to point out that all the initial distributions have the same singularity structure, which is rarely the case for the usual parton sets. Since $\Sigma$ is coupled to the gluon distribution, the latter can also be predicted. We have shown that the fitted gluon distribution is of the same order of magnitude as the gluon distributions obtained by the usual DGLAP fits to DIS data.

By requiring the same singularities in each distribution, we have seen that we are able to construct a full model both for DGLAP evolution and Regge theory in the case of a triple-pole pomeron model. It should be interesting, in the future, to test if we can apply the same method to the case of double pole pomeron or Donnachie-Landshoff two-pomeron model.

In the future, it should also be interesting to see if it is possible to adapt this point of view, in order to derive the triple pole pomeron form factors at high $Q^2$.

\begin{acknowledgments}
I would like to thanks J.-R. Cudell for useful suggestions. This work is supported by the National Fund for Scientific Research (FNRS), Belgium.
\end{acknowledgments}

\end{document}